\documentclass[12pt,preprint]{aastex}

\newcommand{\hst}{{\sl HST}}

\shorttitle{WD Cooling Sequence in NGC 6791}
\shortauthors{Bedin et al.}

\begin{document}

\def\subr #1{_{{\rm #1}}}


\title{Reaching the End of the White Dwarf Cooling Sequence\\
in NGC~6791\footnote{ Based on observations with the NASA/ESA {\it Hubble
Space Telescope}, obtained at the Space Telescope Science Institute,
which is operated by AURA, Inc., under NASA contract NAS 5-26555.}}

\author{ Luigi R.\ Bedin\altaffilmark{2},  
         Ivan R.\ King\altaffilmark{3},
         Jay Anderson\altaffilmark{2,4}, 
         Giampaolo Piotto\altaffilmark{5},  
         Maurizio Salaris\altaffilmark{6}, 
         Santi Cassisi\altaffilmark{7}, and
         Aldo Serenelli\altaffilmark{8}.  
         }

\altaffiltext{2}{Space Telescope Science Institute, 3800 San Martin 
Drive, Baltimore, MD 21218; bedin@stsci.edu}

\altaffiltext{3}{Department of Astronomy, University of Washington,
Box 351580, Seattle, WA 98195-1580; king@astro.washington.edu}

\altaffiltext{4}{Department of Physics and Astronomy, MS-108,
Rice University, 6100 Main Street, Houston, TX 77005;
jayander@stsci.edu}

\altaffiltext{5}{Dipartimento di Astronomia, Universit\`a di Padova,
Vicolo dell'Osservatorio 2, I-35122 Padova, Italy;
giampaolo.piotto@unipd.it}

\altaffiltext{6}{Astrophysics Research Institute, Liverpool John Moores
University, 12 Quays House, Birkenhead, CH41 1LD, UK; ms@astro.livjm.ac.uk}

\altaffiltext{7}{INAF-Osservatorio Astronomico di Collurania,
via M. Maggini, 64100 Teramo, Italy;
cassisi@oa-teramo.inaf.it}

\altaffiltext{8}{Institute for Advanced Study, Einstein Drive,
Princeton, NJ 08540; aldos@ias.edu}

\begin{abstract}
We present new observations of the white dwarf sequence of the old
open cluster NGC 6791.  The brighter peak previously observed in the
white dwarf luminosity function (WDLF) is now better delineated, and
the second, fainter peak that we suggested earlier is now confirmed.
A careful study suggests that we have reached the end of the white
dwarf sequence.
The WDs that create the two peaks in the WDLF show a 
significant turn to the blue in the color-magnitude diagram.
The discrepancy between the age from the WDs and that from the main
sequence turnoff remains, and we have an additional puzzle in the
second peak in the WDLF.  
Canonical WD models seem to fail --at least at $\sim25\%$-level-- in
reproducing the age of clusters of this metallicity. 
We discuss briefly possible ways of
arriving at a theoretical understanding of the WDLF.
\end{abstract}


%
\section{Introduction}
%

NGC 6791 is a very unusual open cluster.  It is one of the richest (an
unpublished star count by one of us (I.R.K.) shows about 3000 cluster
stars brighter than $B\sim 21$), it is old ($\sim$ 8 Gyr) and very
metal-rich ([Fe/H] $\sim$ +0.4, Gratton et al.\ 2006, Carraro et al.\
2006, Origlia et al.\ 2006).
Furthermore, it is close enough that \hst/ACS imaging can reach to
very faint luminosities.  The original aim of our program (GO 9815 and
10471, P.I.\ King) was to study the bottom of the main sequence (MS;
King et al.\ 2005), and we plan to use the new observations presented
here to study the MS down to the hydrogen-burning limit.  However, our
first look at the color-magnitude diagram had made it clear that the
white-dwarf cooling sequence (WDCS) was even more exciting than the
bottom of the MS (Bedin et al.\ 2005a).  The WDCS presented in that
study suggested an age much smaller than the MS turn-off age.

The purpose of the present paper is to extend the WD sequence to fainter
magnitudes by adding a new set of \hst/ACS images acquired since the
previous result was published.  We will show that our new data set is
deep enough to reveal the second, fainter peak in the WD luminosity
function (LF) that was hinted at by Bedin et al.\ (2005a), and that the
WDCS is truncated just below this second peak.

%
\section{Observations and Measurements}
%

\subsection{Observational data}

The Bedin et al.\ (2005a) study was based on only one epoch of data.
Here we add a
second epoch, and also introduce improvements in the reduction
procedures.

Both epochs were taken with the Wide Field Channel (WFC) of ACS.  The
first-epoch images (GO 9815) were taken on 16 July 2003, and consisted
of the following exposures:\ F606W, 4$\times$1185 s + 2$\times$1142 s
+ 3$\times$30 s; F814W, 4$\times$1185 s + 2$\times$1142 s +
3$\times$30 s.  Our second-epoch exposures (GO 10471), which were
taken on 13 July 2005, had been planned before we realized how
interesting and accessible the WDCS would be; as a result our program
was aimed at reaching as faint as
we could on the MS.  We therefore chose to repeat our observations
only in the F814W filter, in order to improve our photometric precision
for the lower MS stars and to obtain proper motions to separate them
from the field stars.  The GO 10471 exposures were 4$\times$1264
s + 2$\times$1200 s + 3$\times$30 s.  All our exposures (in both
epochs) were well dithered.

To strengthen our measurements of the brighter stars, we also measured
archival images taken in GO 10265 (P.I.\ T.\ Brown); these consisted
of \hst/ACS exposures, in each of F606W and F814W, of 0.5, 5, and 50
s, without dithering.  These images overlap about half of our field.
Since their role is only to extend our sparse sampling of the upper
parts of the color-magnitude diagram (CMD), they were not used for the
WDs, but only to study the MS turn-off.

\subsection{Measurements and reduction}

Since our two previous papers on NGC 6791 (Bedin et al.\ 2005a for the
WDCS and King et al.\ 2005 for the MS), our methods of measurement and
reduction have evolved and improved.  The software program used to find
and measure stars is described in detail in Anderson et al.\ (2008).
Here we summarize briefly what the program does.

The first step of all is to construct for each filter a 9$\times$10
array of PSFs that correctly represent the spatial variation of the
ACS/WFC PSF, along with, for each individual image, what Anderson \&
King (2006, AK06) call a perturbation PSF for that image.  The
function of the latter is to account for focus changes related to the
particular breathing state of the telescope during that exposure.

The data are reduced in two separate passes.  The first-pass reduction
uses the code described in AK06 to find the relatively bright stars and
measure positions and fluxes for them.  We then use these star lists to
find the 6-parameter linear transformations from the coordinates in any
image (after correction for geometrical distortion) to those in a
distortion-free reference frame, and also to put the magnitudes into a
common zero-point system.  In doing this we use only stars with solid
and consistent detection in all the exposures.  Here, as in the second
pass, all measurements are made on the {\tt flt} images, which have been
flatfielded and bias-subtracted, but have not been resampled.

The procedures of the second pass are quite different from those of the
first pass.  First, we restrict the analysis to one small
25$\times$25-pixel patch of the field at a time, going through
systematically in a mosaic fashion, so as to cover the whole field.
Each patch is specified by the coordinates of its center in the
reference frame, and we use the transformations to identify the location
of the center of the patch in each of the individual exposures, from
which we extract a 25$\times$25 raster centered on this location.  By
``the individual exposures'' we mean all the exposures in either filter.

A sophisticated software routine then uses the rasters from all of the 
individual exposures to find the brightest sources and to measure a single 
position and an F606W and an F814W flux for each one, using all the exposures 
simultaneously.  It then subtracts out the new-found sources from each 
exposure and iterates the finding procedure until no more objects are found. 
To qualify as found, our automated routines required that an object be 
detected in at least 7 out of the 18 images, and in addition, we required 
these objects to be detected  in a minimum of 2 of the 6 F606W and 
2 of the 12 F814W deep images.  

In order to avoid identifying PSF features as stars, we constructed  
a generous upper-limit model of the extended PSF, including the
diffraction spikes, in such a way as to allow everywhere for the largest
upward excursions that each PSF pixel might reasonably make, and we
insisted that any faint star in the vicinity of a bright star must be 
brighter than any PSF feature that could be present.  This procedure 
does an excellent job of keeping PSF features out of our lists, and 
ends up excluding very few legitimate stars.  

In addition to solving for positions and fluxes during the
simultaneous-fitting process above, we also compute a very important
image-shape parameter for each source:\ RADXS.  This is a measure of
how much flux there is in the pixels just outside of the core, in
excess of the prediction from the PSF.  (We measure it using the
pixels between $r=1.0$ and $r=2.5$, and it is reported relative to the
star's total flux.)  RADXS is positive if the object is broader than
the PSF, and negative if it is sharper.  This quantity is of great
importance in distinguishing between stars and galaxies whose images
are nearly as sharp; the latter are especially numerous in the part of
the CMD where the faint white dwarfs lie.  We experimented with
several similar diagnostic parameters and found that RADXS effected
the best star-galaxy separation for this data set.

For each patch, we compute a stacked representation of the scene in a 
manner similar to Drizzle (Fruchter \& Hook 1992) with {\tt pixfrac} $=$ 0. 
This stack is free of cosmic-rays and image artifacts.  We use this stack to 
compute two other important parameters for each detected object; these are 
the local sky background (SKY) and its r.m.s.\ deviation (rmsSKY).  These
parameters are computed by taking the stack pixels in the annulus between
3.5 and 8.0 pixels from the central pixel of the star and calculating
their mean and the root-mean-square deviation from the mean.  There
is no sigma-clipping done, since we want to be sensitive to all the 
background noise that is present.  The need for the SKY value is
obvious; as for rmsSKY, we will explain below how it gives us a useful
indication of how difficult the local background may make it to measure
faint stars.

Artificial-star (AS) tests are done using a similar patch-based
procedure.  For each AS, a position and an F606W magnitude are chosen
in a random way; the F814W magnitude is then chosen so that the star's
color puts it on the ridge line of the WD sequence.
We next extract from each exposure a patch centered on the AS
location.  The AS is added into the raster for each exposure at the
appropriate location, in the form of an appropriately scaled PSF with
Poisson noise.  The software routine then operates blindly on the
patch, finding and measuring all the stars.  We examine the resulting
list of sources to see if the artificial star was recovered.  The
artificial stars are used, it should be noted, not only as a measure
of completeness; they also serve, at a number of stages of the
procedure, to help develop and calibrate our criteria for choice of
valid stars.

\subsection{Photometric zero points}

We calibrate the photometry to the WFC/ACS Vega-mag system following
the procedure given in Bedin et al.\ (2005b), and using the encircled
energy and the zero points given by Sirianni et al.\ (2005).  Given
the high background ($\sim65$ $e^{-}$/pixel) in these exposures, we
would not expect much of a CTE correction to be necessary.
Nonetheless we did compare the photometry of the two epochs; since
CTE increases linearly with time (Riess \& Mack 2004), any CTE losses
should be about twice as large in the second epoch as in the first.
We plotted $m_2 - m_1$ against the $y$ coordinate and saw no trends,
indicating that any CTE present is much smaller than the random
photometric errors.

\section{Selection of Stars}

By using stacked images to examine stars from our lists, we found that
our automated finding algorithm had successfully identified all of the
faint stars that we might find by eye, and it had also successfully
avoided identifying PSF features as stars.  But the automated
procedure had included in its list many barely resolved galaxies.

\begin{figure}
\epsscale{1.00}
\plotone{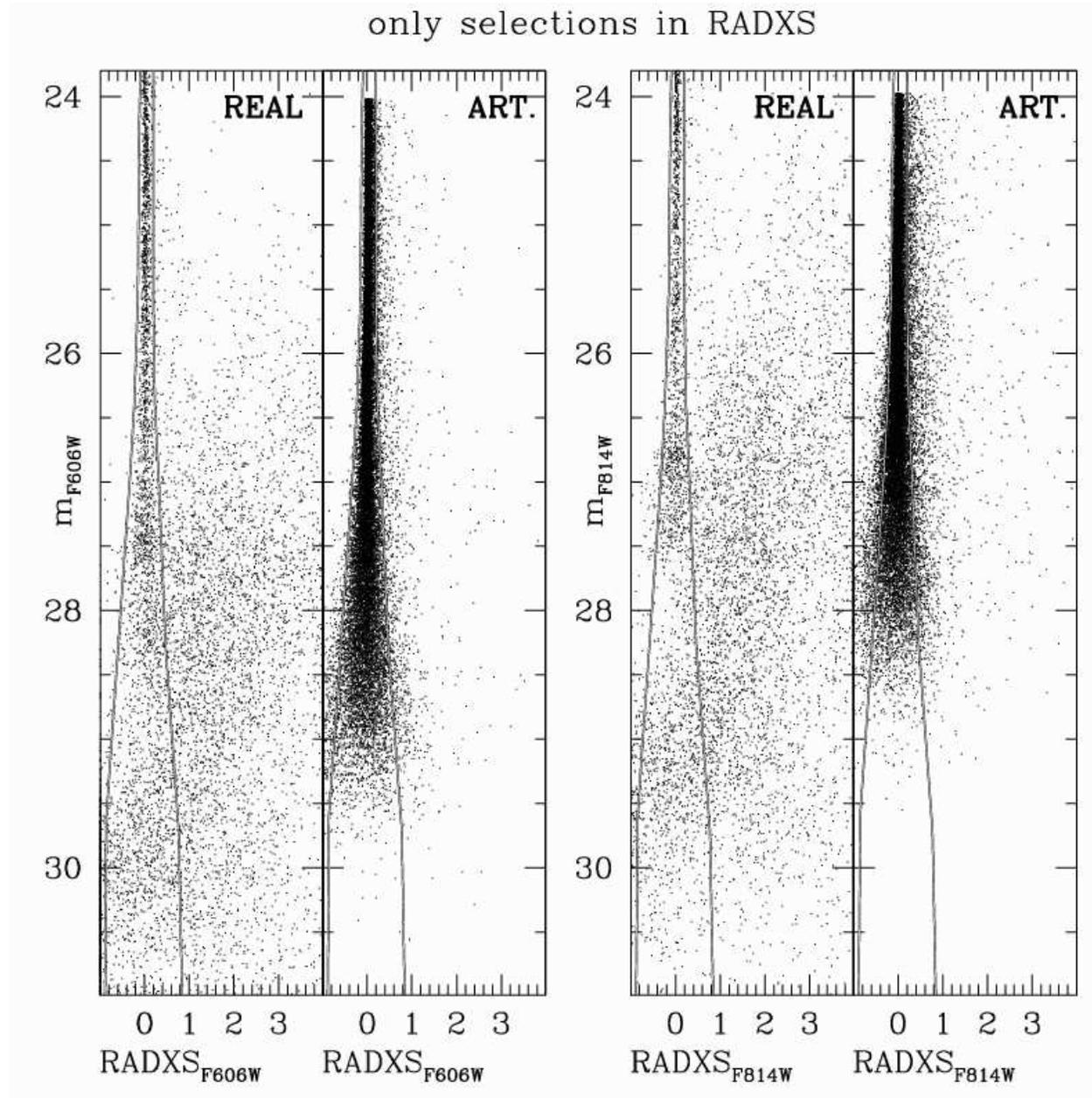}
\caption{ The parameter RADXS as a function of magnitude for artificial
stars and real detections, in filter F606W (left) and in filter F814W
(right). Detections between the two red lines are considered real stars.
Note how the real stars peter out at a much brighter magnitude than the
artificial ones (particularly in the F814W filter).  }
\label{radxs}
\end{figure}

It is particularly important to identify these galaxies, because they
fall in the part of the CMD where the faint end of the WD sequence lies.
To distinguish these barely resolved objects from point sources, we use
the shape parameter RADXS, which measures any excess flux just outside
of the PSF core.  Here the artificial stars are very instructive, since
they show us how RADXS should behave with magnitude for true point
sources.  Figure \ref{radxs} shows us the trend of RADXS against
magnitude (for each filter) for artificial stars on the right of each
panel, and on the left for stars that we consider (thus far) to be real.
It is clear that many of the objects that we have thus far included as
stars are far more extended than are the artificial stars, which we know
have truly stellar profiles because we made them that way.  We use the
distributions for the artificial stars to mark out boundaries in these
diagrams that should retain nearly all the objects that are truly stars;
they are indicated by the red lines, which were drawn in each AS panel
in such a way as to include almost all of the recovered stars; they are
repeated identically in the corresponding real-star panel.  (We believe
that the tail-off of objects to the right in the AS panels is produced
by star-star blends, which certainly should be eliminated from our
photometry.)

\begin{figure}
\epsscale{1.00}
\plotone{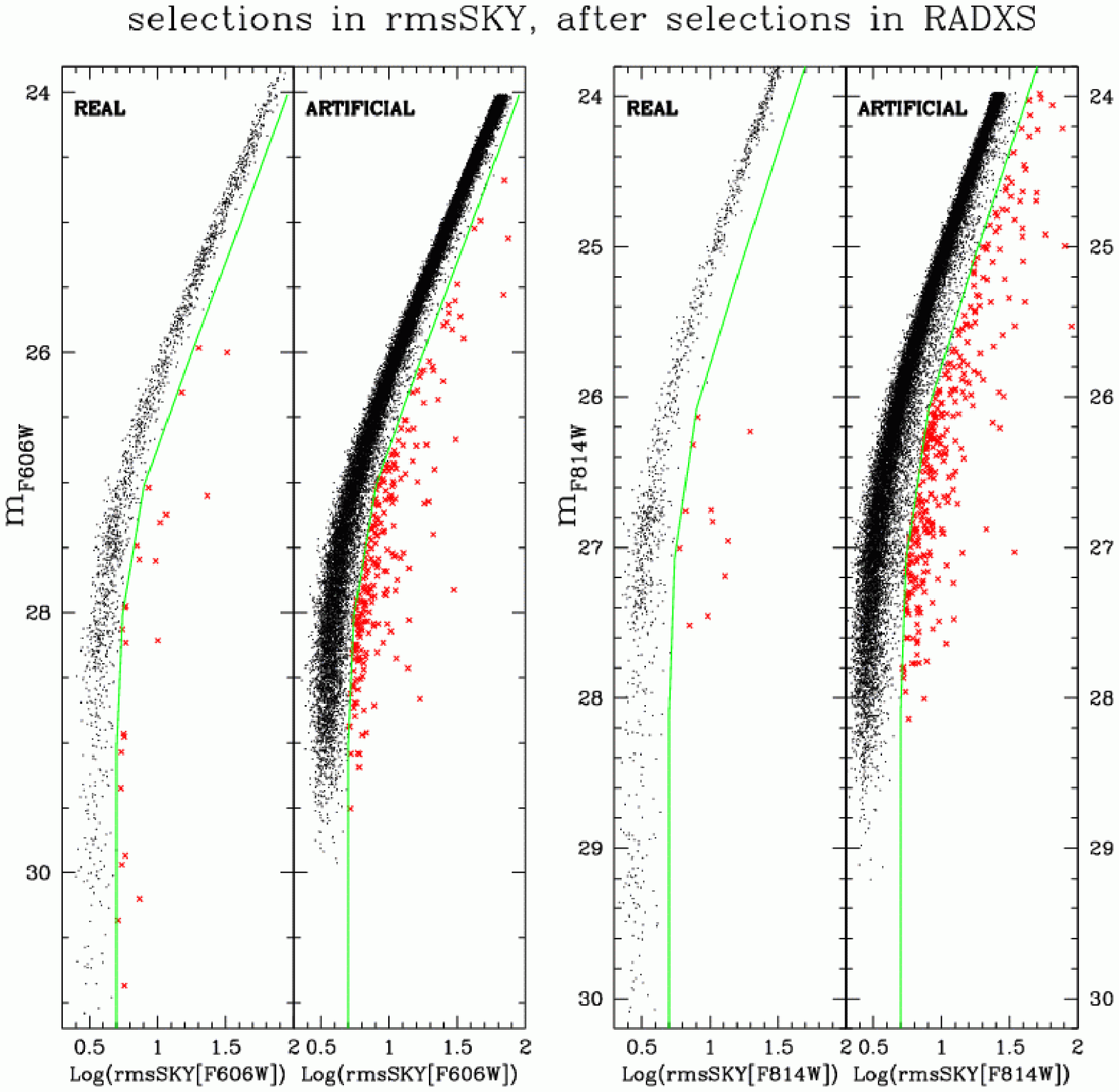}
\caption{For stars that passed the RADXS selection, we show here the
         log$_{10}$ of the rmsSKY parameter as a function of magnitude, 
         for both artificial stars and real detections, in filter F606W 
         (left) and in filter F814W (right). Stars with rmsSKY to the 
         left of the thin line are likely to be well measured.  The 
         others, plotted as crosses, are rejected.  }
\label{rmsSKY}
\end{figure}

Our second image-quality parameter, rmsSKY, hardly figures in the direct
selection of stars, but it does tell us how suitable the surroundings of
each star are for good photometry.  An unusually high value of rmsSKY
indicates a region where the background is irregular --- often caused by
the mottled halo of a bright star's PSF.  We say ``unusually high''
rather than just ``high'' because even in the absence of interfering
objects, rmsSKY will increase with the brightness of the star around
which it is measured; because it must be measured only a few pixels from
the star, its value has an increase that comes from the PSF wings of the
star image around which it is measured.  The value of SKY in the
measuring annulus goes up, and its Poisson noise increases the value of
rmsSKY.  This effect is clearly seen in Figure \ref{rmsSKY}, which displays 
rmsSKY for real and artificial stars in the two filters in a way exactly
analogous to the way Fig.~\ref{radxs} displayed RADXS.  Again, limit lines have
been drawn by eye for the artificial stars, in a way that distinguishes
the large majority of normal cases from the small proportion of
disturbed ones.  The same limiting lines, when applied to the real
stars, served to eliminate those stars whose photometry is suspect
because of irregularity of their background.  We will use rmsSKY again
in Section \ref{compl}, where it plays an even more important role.

\begin{figure}
\epsscale{1.00}
\plotone{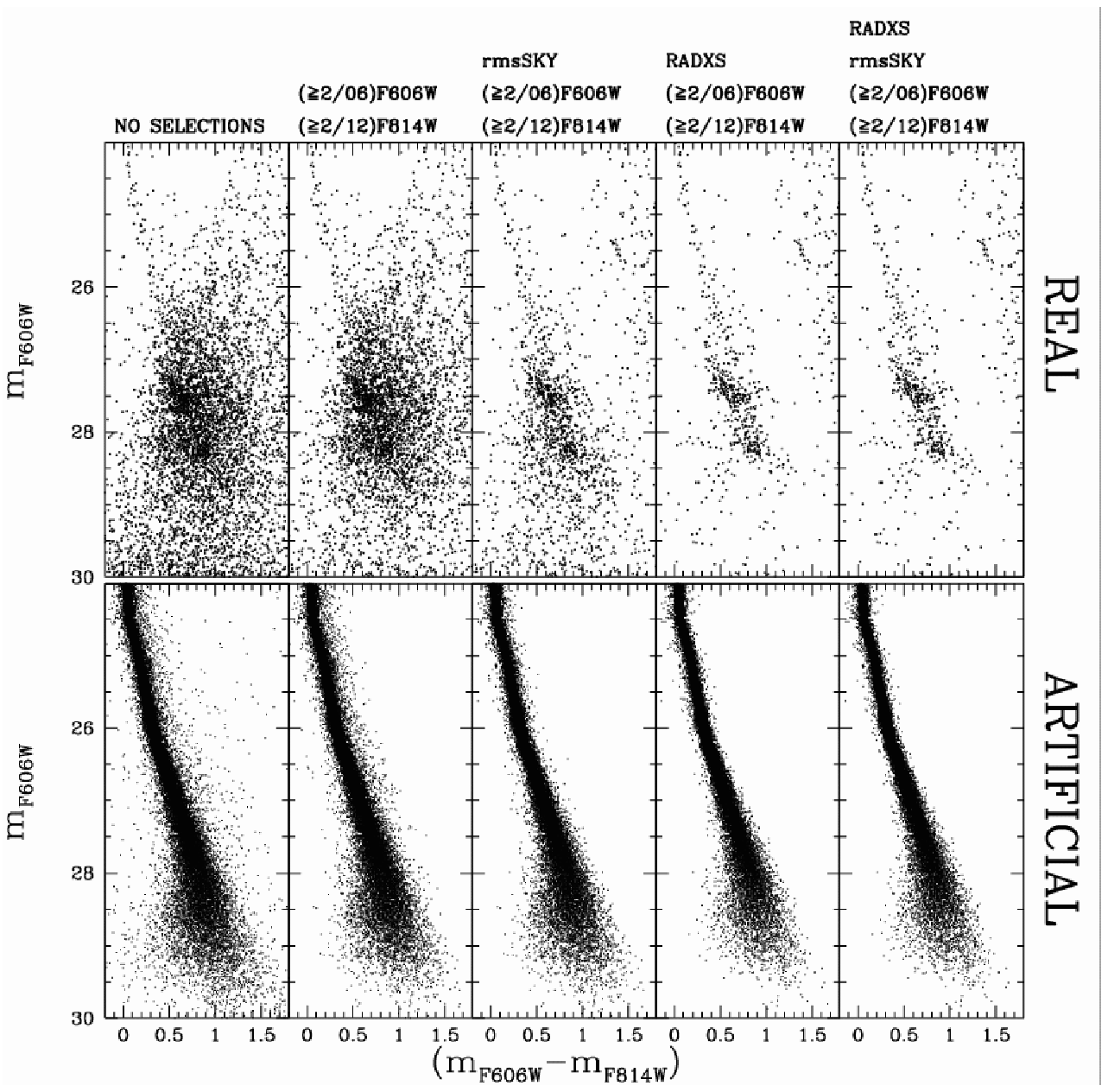}
\caption{{\em(Top):} The CMD of NGC 6791 with various selection
         criteria applied to the individual stars. 
         {\em(Bottom):} The same selection applied to the recovered 
         artificial stars.}
\label{cmds}
\end{figure}

Figure \ref{cmds} shows the WD region of the CMD of NGC 6791, with
various selection criteria:\ first, all objects that were measured at
all, then all stars that were measured in at least two deep images for
each filter, next the survivors when only the rmsSKY criterion is used
for selection, then all the objects that met the RADXS criterion, and
finally the stars that qualified with respect to both RADXS and rmsSKY.
Since the last two panels are nearly identical, it is clear that rmsSKY
removes very few objects that were not already removed by RADXS.

The WDCS that we find here for NGC 6791 is consistent with the one
exhibited by Bedin et al.\ (2005a), and the second peak, which was
hinted at in their diagram, is now quite clear.

\section{Completeness}
\label{compl}

The artificial stars were carried through all the stages of the
measurement procedure in exactly the same way as the real stars, so as
a measure of completeness we simply counted the fraction of artificial
stars that were recovered, as a function of magnitude.  We considered
an AS to be recovered if its measured position was within 0.75 pixel
and its magnitude within 0.5 mag of the input values, for both
filters.

This is a simple and straightforward approach to completeness, and it
does indeed provide the correct factor, as a function of magnitude, by
which to divide the observed numbers in order to derive
completeness-corrected numbers.  
There is great value, however, in looking at completeness in a 
more detailed way.  For faint stars our numbers are less complete for two 
reasons.  One is simply the difficulty of detecting faint stars.  Even 
under the best of circumstances we will lose some fraction of the stars 
on account of statistical fluctuations that work against their detection.
But a great deal of our loss of faint stars is for a quite different 
reason:\ there are regions of the image in which conditions are very 
unsuitable for the detection of faint stars; there is a quite appreciable 
fraction of the image area, in fact, in which even many of the bright 
stars are lost, because of interference by even brighter stars.  

The distinction between these two possible reasons for missing stars
becomes important when we consider the other application of the curve of
completeness against magnitude:\ as a criterion of the reliability of
the number of stars.  It is customary to consider star numbers to be
unreliable when the completeness figure has dropped below 50\%.  But
which completeness?  Imagine a field in which 60\% of the total area is
unsuitable for finding faint stars at all; the conventional completeness
cannot be above 40\% at any faint magnitude, and judging reliability
from it would lead to the absurdity that there is no way to say anything
about the faint population.  Yet if we were to confine our measurements
to that 40\% of the field in which faint stars can be measured, we would
correctly conclude that the star counts are reliable down to a fairly
faint level.

Drawing the map of reliable areas --- probably a quite complicated map
--- would be a daunting task, and this is where our rmsSKY parameter
becomes so useful.  The conventional completeness is $c =
N\subr{rec}/N\subr{ins}$, where $N\subr{ins}$ and $N\subr{rec}$ are the
numbers of stars inserted and recovered, respectively.  But we can also
define a good-region
completeness, $c\subr{g} = N\subr{rec,g}/N\subr{ins,g}$; this is the
completeness evaluated only in regions of low rmsSKY, where stars of
that magnitude
could be found without interference.  The limit of reliability of our
numbers is at the 50\% level of $c\subr{g}$, not of $c$.

\begin{figure}
\epsscale{1.00}
\plotone{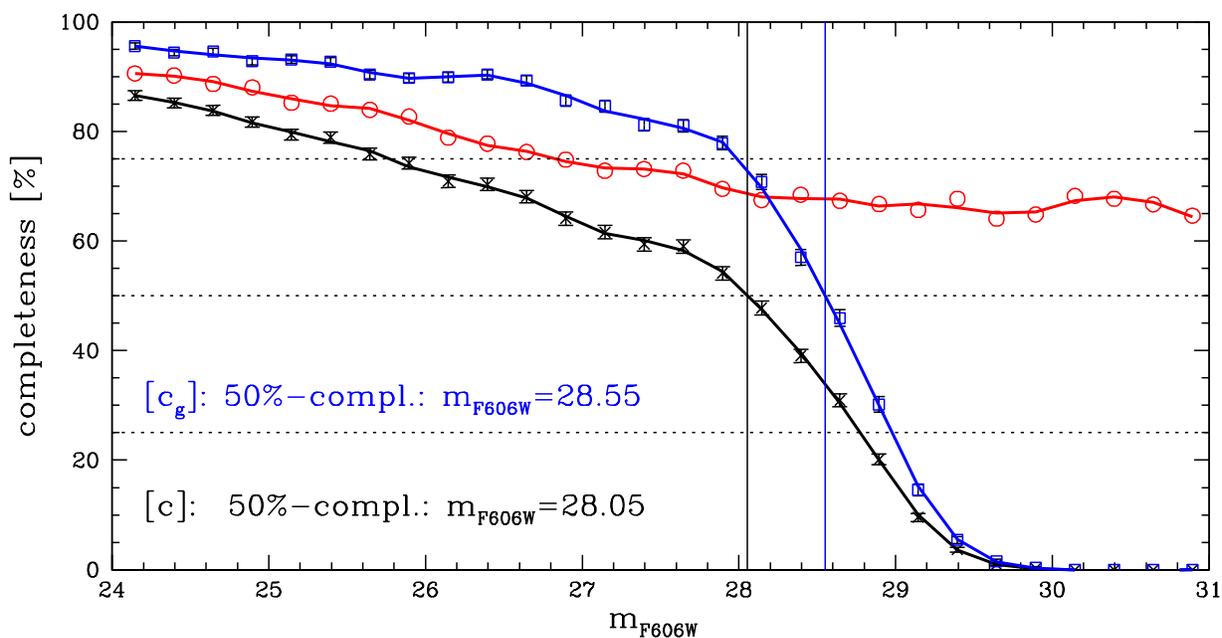}
\caption{The crosses show the conventional completeness $c$, while the
         squares are the low-rmsSKY completeness $c\subr{g}$ that is 
         defined in the text. The circles indicate the fraction of the 
         image area where the bumpiness of the of the sky offers no 
         impediment to finding a star, at each magnitude.  The product of 
         this curve
         and $c\subr{g}$ is of course $c$.  }
\label{Comp_g}
\end{figure}

Figure \ref{Comp_g} shows the completeness levels for our field.  
The 50\% level of
$c$ is at F606W magnitude 28.05, while for $c\subr{g}$ it is at 28.55.
If we had followed the absurdity of using the 50\% level of conventional
completeness, instead of the informed choice of the 50\% level of
$c\subr{g}$, we would have lost the interval of half a magnitude that
makes the difference between going deeper than the lower cutoff of the
WD CS and failing to do so.
Figure \ref{Comp_g} also shows the fraction of the image area suitable
for detection of faint sources (open circles).

\section{Proper motions}

Here, to compute a proper motion for each star, we first measured a mean
position for the star in the reference frame at each epoch, using all
the exposures within that epoch.  The proper motion, then, is the
difference between the second- and first-epoch positions, divided by
the baseline, and multiplied by the pixel scale (50 mas/pixel).  The
reference frame is aligned to have $x$ and $y$ parallel to RA and Dec,
respectively, so the proper motions are in a properly aligned frame as
well.  The zero point of the motion is the cluster's bulk motion,
since the reference frame was defined by member stars.

\begin{figure}
\epsscale{1.00}
\plotone{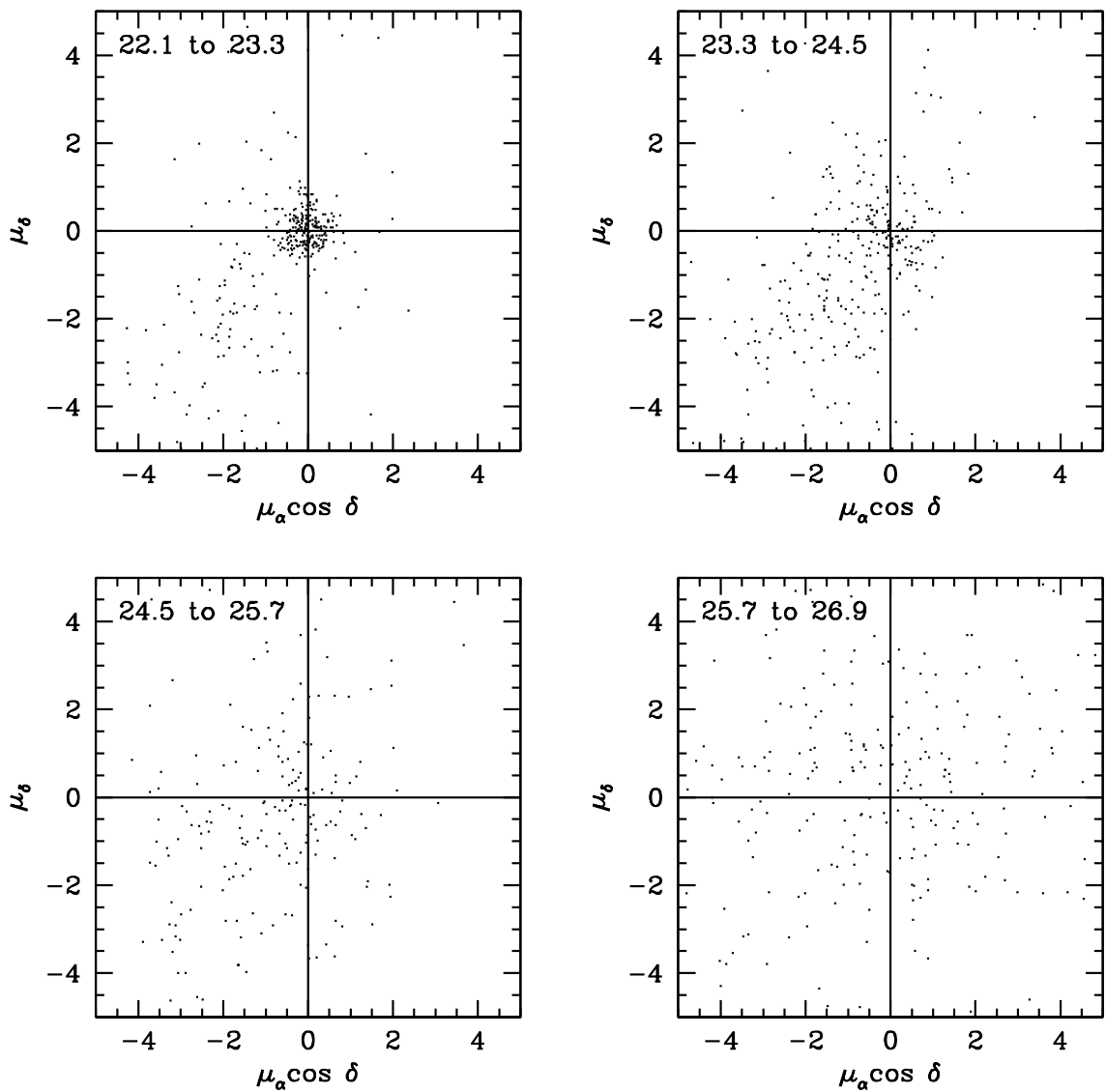}
\caption{Proper motions for increasingly faint intervals of F814W 
         magnitude.  Zero point is the mean cluster motion.  In the 
         brightest interval the concentration of motions clearly 
         identifies the cluster members, while field stars are more 
         widely spread, and largely in the lower left-hand quadrant.  
         At fainter magnitudes the measuring errors dominate, and no
         separation is possible.  }
\label{pmerr}
\end{figure}

Although our proper motions are very helpful for cleaning field stars
from the upper main-sequence part of the CMD and for the brighter
white dwarfs, for the faint white dwarfs they turn out to be of very
little use.
This can be seen in Figure~\ref{pmerr}.  For the brighter stars (upper
left), the proper-motion precision easily allows us to separate the
two populations.  But as the stars get fainter, the proper motions
become less accurate, the distributions for cluster and field overlap,
and no accurate discrimination is possible.
As is clear from the figure, there is a small difference between the
mean proper motions of cluster and field stars, but by far the best
discriminant is the very small dispersion of the motions of the cluster
stars.  It is equally clear from the figure that a clean separation
would require a very much better accuracy in the motions of the faint
stars --- something that appears to be totally beyond our reach with
the data sets that are available at present. 

\section{The White-Dwarf Luminosity Function}
\label{WDLF}

Without a proper-motion elimination of field stars, the best we can do
toward deriving a white dwarf luminosity function (WDLF) is to use the
RADXS parameter to remove as many galaxies and artifacts as we can,
and delimit in color the region of the CMD in which we consider stars
to be white dwarfs.  For both of these tasks the artificial-star
experiments prove extremely useful.  First, Fig.~\ref{radxs} shows our
treatment of the parameter RADXS.  Independently for each filter, we
used the AS tests to draw our discriminating lines, in such a way as
to include nearly all legitimate objects; we then applied those same
selection criteria to the real stars.  A star was accepted only if it
passed the RADXS test in both filters.

\begin{figure}
\epsscale{1.00}
\plotone{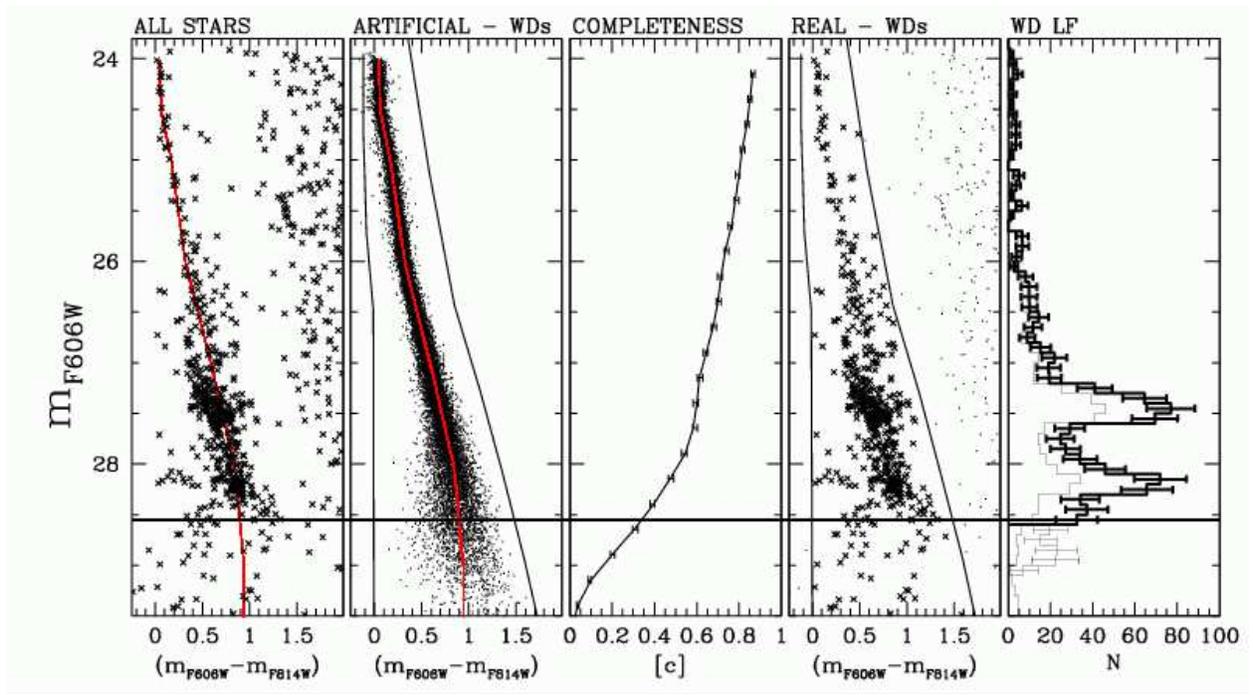}
\caption{The WDLF, including corrections for the completeness curve
         shown in the central panel.  (Successive panels are observed CMD,
         artificial stars, completeness, selection of stars, and LF.)
         The horizontal line at magnitude 28.55 is the 50\%-completeness 
         level of $c\subr{g}$.  }
\label{wdlf}
\end{figure}

We then chose our boundaries in the CMD, as indicated in the second and
fourth panels of Fig.~\ref{wdlf}.  The boundaries were chosen so that
none of the artificial stars were lost, while the boundaries are set
broadly enough to include all stars that might be white dwarfs of some 
sort --- including WD binaries, and unusual types of WDs.

In the middle panel of Fig.~\ref{wdlf} is the completeness.  According
to the distinction made in Section \ref{compl}, we plot here the quantity 
$c$, which is the factor by which observed star numbers need to
be divided in order to estimate the correct total.  The horizontal line
is the 50\% completeness level of the quantity $c\subr{g}$, which, as 
we explained in Section \ref{compl}, is a realistic indicator of our 
ability to find stars in regions where interference does not prevent 
them from being found.

The rightmost panel of Fig.~\ref{wdlf} shows the WDLF.  The observed
numbers are indicated by a thin-line histogram, and the
completeness-corrected numbers are indicated by the thick-line
histogram.  Error bars are shown only on the corrected values, but they
are derived from the Poisson uncertainties in the observed values.  The
corresponding numbers are in Table \ref{WDLFtab}.  The LF below $m_{\rm
F606W}=28.55$ is highly suspect both because of the low completeness and
because the stars in this region show no coherence in color and are
likely not to be WDs at all.  For this reason that part of the LF is
plotted with lines that are less heavy.

\begin{table}[ht!]
\caption{Completeness-corrected white dwarf luminosity function.}
\begin{tabular}[h]{ccrllcrr} 
& & & & & & & \\
$m_{\rm F606W}$ & $N_c$ & $\sigma_{N_c}$ & ~~ & ~~ & 
                $m_{\rm F606W}$ & $N_c\,$ & $\sigma_{N_c}$\\  
& & & & & & & \\
\hline\hline
23.05 &  0.00 &  0.00 &  &  & 26.15 &  8.38 &  3.42 \\
23.15 &  0.00 &  0.00 &  &  & 26.25 &  9.87 &  3.73 \\ 
23.25 &  0.00 &  0.00 &  &  & 26.35 &  9.96 &  3.77 \\ 
23.35 &  1.11 &  1.11 &  &  & 26.45 & 10.07 &  3.81 \\ 
23.45 &  0.00 &  0.00 &  &  & 26.55 & 14.56 &  4.61 \\ 
23.55 &  1.12 &  1.12 &  &  & 26.65 & 11.80 &  4.17 \\ 
23.65 &  0.00 &  0.00 &  &  & 26.75 &  9.03 &  3.69 \\ 
23.75 &  0.00 &  0.00 &  &  & 26.85 & 15.36 &  4.86 \\ 
23.85 &  0.00 &  0.00 &  &  & 26.95 & 21.93 &  5.86 \\ 
23.95 &  1.14 &  1.14 &  &  & 27.05 & 19.17 &  5.53 \\ 
24.05 &  2.30 &  1.63 &  &  & 27.15 & 19.54 &  5.64 \\ 
24.15 &  4.62 &  2.31 &  &  & 27.25 & 41.05 &  8.21 \\ 
24.25 &  1.16 &  1.16 &  &  & 27.35 & 64.56 & 10.34 \\ 
24.35 &  2.34 &  1.65 &  &  & 27.45 & 76.95 & 11.35 \\ 
24.45 &  1.18 &  1.18 &  &  & 27.55 & 69.44 & 10.85 \\ 
24.55 &  2.37 &  1.68 &  &  & 27.65 & 29.18 &  7.08 \\ 
24.65 &  3.59 &  2.07 &  &  & 27.75 & 24.67 &  6.59 \\ 
24.75 &  3.62 &  2.09 &  &  & 27.85 & 27.17 &  7.01 \\ 
24.85 &  3.66 &  2.11 &  &  & 27.95 & 33.99 &  8.01 \\ 
24.95 &  1.23 &  1.23 &  &  & 28.05 & 45.89 &  9.57 \\ 
25.05 &  0.00 &  0.00 &  &  & 28.15 & 71.94 & 12.34 \\ 
25.15 &  5.01 &  2.50 &  &  & 28.25 & 65.72 & 12.20 \\ 
25.25 &  3.79 &  2.19 &  &  & 28.35 & 34.16 &  9.13 \\ 
25.35 &  1.27 &  1.27 &  &  & 28.45 & 37.28 &  9.96 \\ 
25.45 &  6.43 &  2.87 &  &  & 28.55 & 32.46 &  9.79 \\ 
25.55 &  1.30 &  1.30 &  &  &------ &------ &------ \\ 
25.65 &  0.00 &  0.00 &  &  & {\it 28.65} & {\it 19.87} & {\it  8.11} \\ 
25.75 &  6.65 &  2.97 &  &  & {\it 28.75} & {\it 15.36} & {\it  7.68} \\ 
25.85 &  6.75 &  3.02 &  &  & {\it 28.85} & {\it 22.85} & {\it 10.22} \\ 
25.95 &  4.10 &  2.37 &  &  & {\it 28.95} & {\it 22.44} & {\it 11.22} \\ 
26.05 &  2.76 &  1.95 &  &  & {\it 29.05} & {\it  7.22} & {\it  7.22} \\
\hline 
\hline
\label{WDLFtab}
\end{tabular}
\end{table}

The WDLF obtained here very much resembles the one previously obtained
by Bedin et al.\ (2005a, top panel of their Fig.~3), but we now see
features that are somewhat better defined, as a result of deeper
integration in F814W and a more careful sample selection.  The WDLF
shows the expected mild rise down to $m_{\rm F606W}\sim 27$, but then
come the two unexpected peaks.  (i) We confirm the presence of the first
peak at a magnitude $m_{\rm F606W}=27.45\pm0.05$, better constrained
than in our previous paper, and with a steeper drop than its rise ---
such as we might expect at the end of a cooling sequence.  (ii) But then
there is a clear second peak, at magnitude $m_{\rm F606W}=28.15\pm0.05$.

We believe that the WDLF ends at a magnitude where our completeness is
still reliable.  This conclusion is strengthened by the CMD shown in the
first and fourth panels of Fig.~\ref{wdlf}, where any continuation of
the LF is contributed by stars whose color is off that of the ridge-line
of the WD sequence, so that they may well not be cluster white dwarfs at
all.  AS tests indicate that if the WDCS continued below $m_{\rm
F606W}=28.4$, we would see a well-defined sequence, as opposed to the
sparse and scattered cloud that we do see, which certainly cannot
correspond to a continuing sequence.

\begin{figure}
\epsscale{1.00}
\plotone{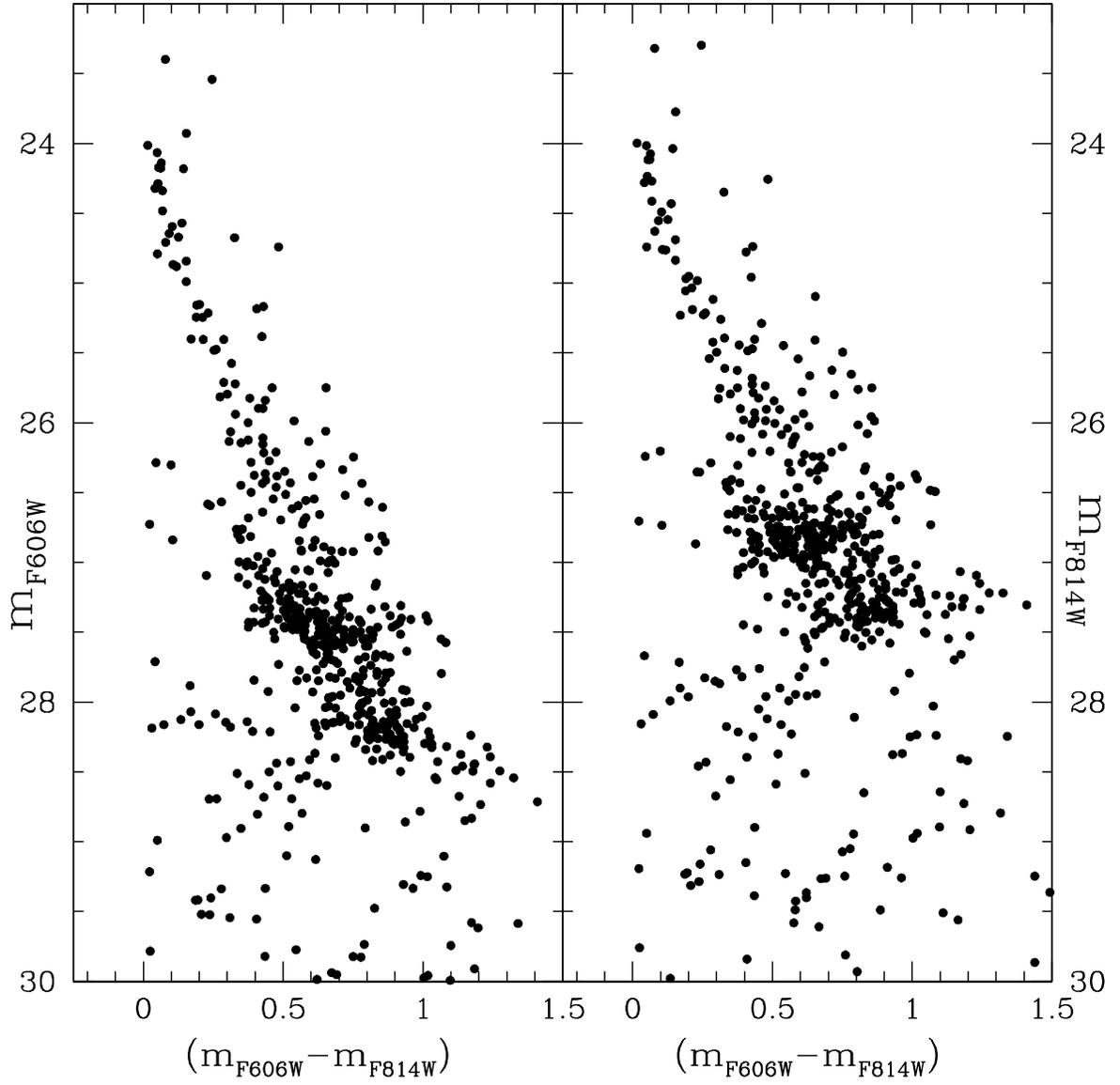}
\caption{Color-magnitude diagrams of the WDCS.}
\label{WDs}
\end{figure}

Finally, we note that at the magnitude levels of the two LF peaks, the
WDs stand on the blue side of the cooling sequence in the CMD, forming
a sort of blue hook.  This feature is better seen in 
Fig.~\ref{WDs}, where we zoom around the
fainter part of the WDCS, and it is expected from models (see next
section).  Note that the two blue hooks create the LF peaks.

At the time of writing Bedin et al.\ (2005a), we estimated that the
total number of observed WDs was consistent with the number to be
expected from a Salpeter IMF normalized to the present-day MF for MS
stars (which we took from King et al.\ 2005).  We can now re-examine
this question.  Down to magnitude $m_{\rm F606W}\sim28.55$ we now
observe 425 WDs that we assume to be NGC 6791 members; correction for
incompleteness brings this to $\sim850$.  (We already commented that we
do expect very few, if any, WDs below $m_{\rm F606W}\sim28.55$.)  By
contrast, the number of WDs expected from a Salpeter IMF normalized to
the present day MF is $\sim430$.  Using a Kroupa (2001) IMF we would
expect $\sim$500 WDs, still a factor of about two less than the measured
counts.  

We must treat these numbers with great reserve, however.  
They include no correction for background galaxies that may have been
included in our WD count, though we expect it to be small because of
the care we took in eliminating non-stellar objects (see Section 3). 
We will return in Sect.~\ref{hudf} to the question of contamination by
background objects. 
As for the predicted number of WDs, it was calculated without taking
into account the fact that our field lies completely within the core
of the cluster, and it does not take mass-segregation effects into
account.  Also, the assumption of a single slope for the power-law
mass function (or of a Kroupa 2001 MF) might be an oversimplification.

\subsection{Spatial distribution of the WDs}
\label{xywd}

\begin{figure}[ht!]
\epsscale{1.00}
\plotone{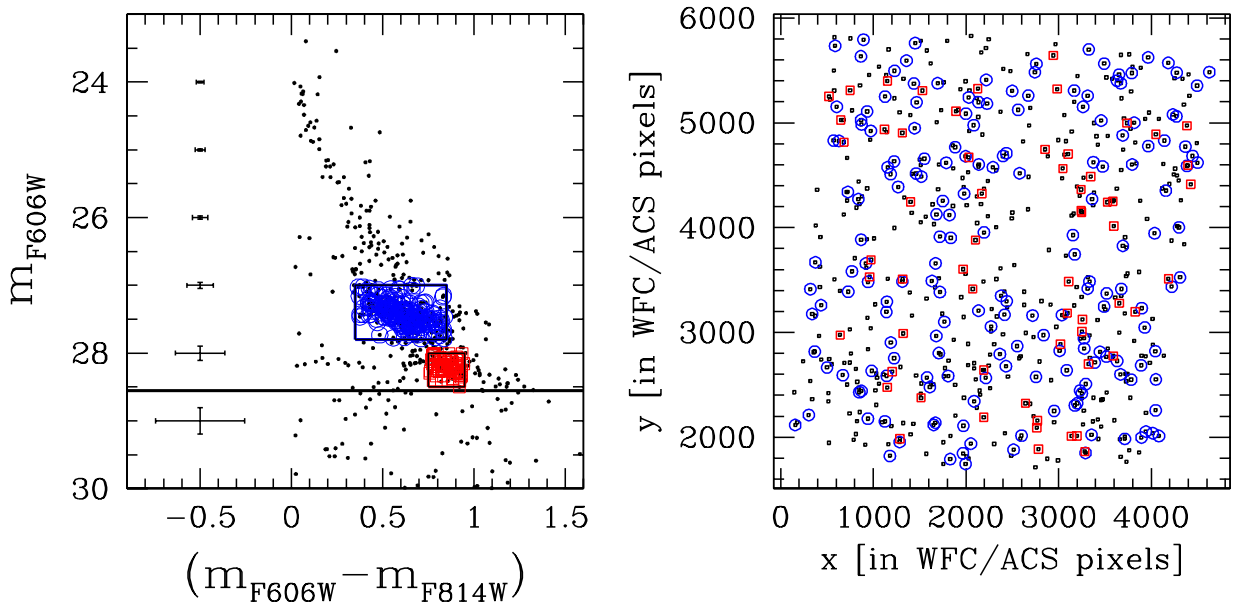}
\caption{({\it left}) CMD of the WD region, with the completeness limit
marked.  The error bars show the photometric errors in magnitude and
color, as a function of magnitude.  The two boxes delimit the areas that
we took as representative of the two LF peaks.  ({\it right}) Spatial
distributions of objects.  The black dots are all those above the
completeness limit.  On both sides objects belonging to the brighter
peak are marked in blue (circles) and those belonging to
the fainter peak in red (squares).}
\label{xy}
\end{figure}

At the suggestion of the referee, we exhibit in this section the spatial
distribution of the WDs in the two peaks of the LF.  On the left side of
Fig.~\ref{xy} we plot the WD CMDs and delineate with boxes the blue-hook
regions that contain nearly all of the stars in the two peaks.  The
horizontal line marks the 50\% completeness limit for regions of low
rmsSKY, as discussed in \S\ \ref{compl}.  The right side of the figure
shows the spatial distribution of all the objects above this
completeness limit, with blue circles around the dots that represent
stars in the upper rectangle and red squares for those in the lower
rectangle.  No differences are evident between the distributions of the
red, the blue, and the plain dots.

\section{Contamination by background galaxies}
\label{hudf}

There is clearly a danger that our supposed WD sample includes a number
of sharp nuclei of background galaxies, as these objects have colors
close to those of the WDs.  Here we attempt to address this question.
Following a suggestion by the referee, we have chosen a subset of the
images in the Hubble Ultra Deep Field (HUDF, Beckwith et al.\ 2006),
in which practically all detected objects should be galaxies.
The suggestion was that we choose a set of images with a total exposure
similar to what we had in NGC 6791, and measure them in the same way.

Unfortunately the HUDF observations did not include the F814W filter;
the closest filters to it were F775W and F850LP.  We chose to work
with F775W, because it is easier to transform into the F814W band.
The throughputs of F775W and F814W are close enough that we felt that
we could simply match the total exposure times.  The HUDF images used
here thus consist of 6$\times$1200 s exposures in F606W, and
12$\times$1200s in filter F775W.

The reduction, selection of objects, calibration, and analysis were
done in exactly the ways described in the previous sections, except
that we used unperturbed library PSFs.  The RADXS parameter seemed to
do an extremely good job of rejecting the non-point-like sources.  We
did not make artificial-star runs on the HUDF images, however, because
we did not consider it to be necessary in this rough exercise.  We
simply assume that because the total exposure was chosen to be
similar, objects that are at the completeness limit in (F606W, F814W
will also be close to the completeness limit in (F606W, F775W).

\begin{figure}
\epsscale{1.00}
\plotone{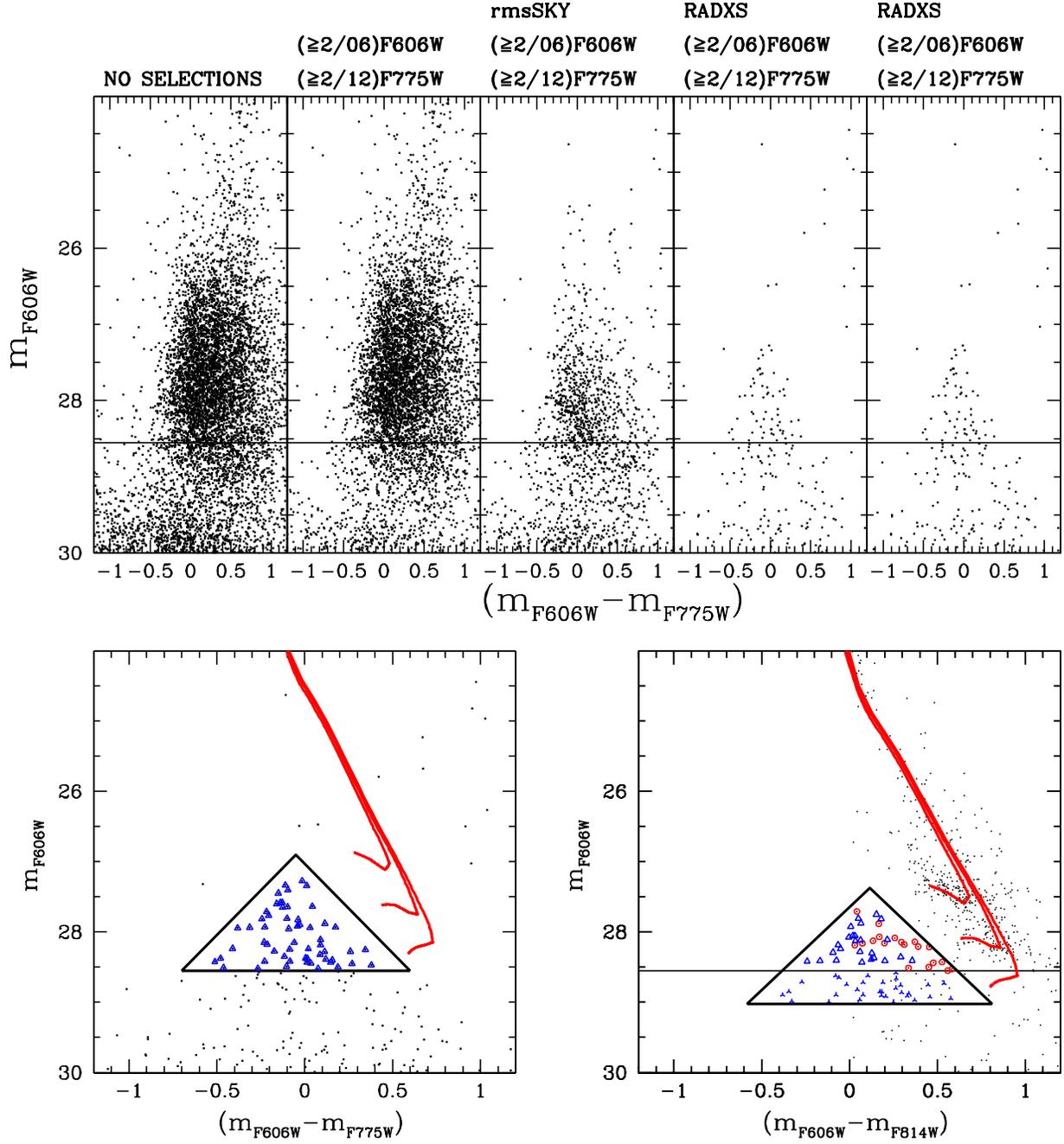}
\caption{({\it top}) As in top panel of Fig.~\ref{cmds}, but for the
  HUDF.
  ({\it bottom left}) Faint part of WD region of CMD, in HUDF, with
  F775W for $I$ filter.  ({\it bottom right}) Same, but for NGC 6791, with
  F814W for $I$ filter.  For the meaning of the lines, triangles, and
  highlighted stars, see text.}
\label{udf}
\end{figure}

In the HUDF essentially all objects are galaxies.  In the upper-part of
Fig.~\ref{udf} we present (on a more compressed color scale) CMDs for
the the objects in the HUDF across all the various selection stages,
exactly as in the top panels of Fig.~\ref{cmds}, but with the F775W
filter instead of F814W.  Note how the galaxies that have slipped
through our selection process are nearly all confined to a narrow
triangular plume extending up from our completeness limit.

In the lower half of the figure is shown a portion of the CMD.  On the
left is the HUDF CMD in its color system, on the right the NGC 6791
CMD in its color system.  We drew in the HUDF CMD a triangle extending
downward to an arbitrary but generously faint baseline; it is shown by
heavy lines.  The blue symbols (small-triangles) within the triangle
represent the HUDF objects that met all our stellar criteria.
Adjacent to the right edge of the triangle are theoretical WD
isochrones (which will be discussed in \S\ \ref{isochr}) in the color
system of this CMD.  Since the later discussion will show that they
fall in the region of the WDCS, it is clear that our triangle falls
close to where the WD sequence should lie.

We then transformed the vertices of this triangle into the magnitude
system of the NGC 6791 CMD.  To do this we used the transformations from
\hst\ filters to Johnson-Cousins ($V$, $I$) given by Sirianni et al.\
(2005); next we used the Sirianni relations to transform from ($V$, $I$)
to (F606W, F814W).  (Note: As the Sirianni procedures are set up, the
transformation from \hst\ filters to Johnson-Cousins requires an
iterative process, whereas the transformation back to a \hst\ system
requires only a direct evaluation.)  The triangle as transformed into
the NGC 6791 system also required applying the interstellar absorption,
for which we assumed 0.47 mag in F606W and 0.31 mag in F814W, in
agreement with the absorption and reddening used elsewhere in this
paper.  The same isochrones are repeated here, but in the color system
of this CMD and with the same adjustments for absorption.

We transformed not only the vertices of the triangle, but also the
HUDF objects that lay within it.  We then truncated this triangle at
our completeness limit, $m\subr{F606W}=28.55$.  Within the remaining
area, the lower right panel of the figure shows the triangle, and
within it the NGC 6791 stars in red and the transformed HUDF objects
in blue.  (The HUDF objects below the completeness limit are plotted
as tri-pointed asterisks; they were excluded from the discussion that
follows.)

One conspicuous difference is that the HUDF objects extend to a
noticeably bluer color.  In trying to understand this, we note that the
HUDF objects are non-stellar, whereas we have used color transformations
that are appropriate only for stars.  Since we do not know the spectral
energy distributions (SEDs) of these objects, the difference could be
due to having used stellar SEDs rather than the correct but unknown
ones.  
Because the NGC 6791 objects extend continuously across the right edge
of the triangle as it is shown, the number of them that fall within the
triangle is obviously quite sensitive to the color transformation, and
is therefore rather uncertain.
What the lower right panel does make clear, though, is that it is
highly likely that a significant number of NGC 6791 objects on the
blue side of the WDCS are likely to be the nuclei of background
galaxies.

In any case, and most importantly, we note that the number of objects
in question is small --- at most a few dozen.  This means that this
source of possible contamination cannot be responsible for the second
peak in the WD LF, nor for the discrepancy between the predicted and
observed numbers of WDs that was noted in \S\ \ref{WDLF}.

\section{Theoretical Implications}

Our new luminosity function of the white dwarfs in NGC 6791 confronts us
with two theoretical problems:

\begin{enumerate}

\item The LF has two distinct peaks, rather than a single one.

\item Fitting either peak with a theoretical LF calculated from
  conventional WD cooling theory leads to a WD age that is smaller
  than the age derived from the MSTO.  (We showed this for the brighter
  peak in Bedin et al.\ 2005a.  We will confirm this discrepancy below,
  and will show that it exists even if we fit the fainter peak instead.)

\end{enumerate}

In this section we will explore ways in which conventional theory might
be modified in order to cope with these two problems.

\subsection{Fitting with the conventional theory}
\label{isochr}

We have calculated WD isochrones and LFs with [Fe/H] $=$ 0.40 for the
WD progenitors (following the most recent spectroscopic estimates by
Gratton et al.\ 2006, Carraro et al.\ 2006, and Origlia et al.\ 2006).
All models and isochrones are from the BaSTI database (Pietrinferni et
al.\ 2004), and the Salaris et al.\ (2000) WD tracks have been
transformed into the ACS/WFC Vega-mag system as described in Bedin et
al.\ (2005b).
The adopted CO profiles of our WD models are from Salaris et al.\ (1997).
The initial-final mass relationship is from Ferrario et al.\ (2005),
extrapolated linearly for initial masses below $2.5M_{\odot}$ (i.e.,
below the lower $M\subr{i}$ limit of the Ferrario et al.\
determination).  The relationship implies that $M\subr{f}=0.53M_{\odot}$
at $M\subr{i}=1.0M_{\odot}$.  
As a test, we also examined the different $M\subr{i}$--$M\subr{f}$
relationship used in the Richer et al.\ (2000) WD isochrones, and in
Kalirai et al.\ (2007), but the results we obtain are virtually the
same. 

The [Fe/H] $=$ 0.40 isochrones have been used to fit our \hst\
results; we derived $E(B-V)=0.17$, $(m-M)_V =13.50$, and a turn-off
(TO) age of about 8~Gyr.  A fit to our \hst\ CMD employing these
values for reddening and distance modulus is displayed in the upper
panels of Fig.~\ref{fit}, for both the TO and the WDs.  For reference,
this distance modulus requires an age of $\sim6$~Gyr (about 2~Gyr
younger than the TO age) to reproduce the WDLF peak at $m_{\rm
F606W}\sim28.2$, or an age of $\sim4$~Gyr ($\sim4$~Gyr younger than
the TO age) to reproduce the peak at $m_{\rm F606W}\sim27.4$, as also
shown in the lower panel of Fig.~\ref{fit}.

Even though the ages associated with these fits are presumably wrong, we
note that each of them does reproduce the observed hook toward the blue
and toward brighter magnitudes.  This hook is caused by the more massive
objects (coming from higher-mass progenitors) that pile up at the bottom
of the WD isochrone, moving the track in the blue direction because
these higher-mass WDs have smaller radii.  It is important to notice
that this blue hook has nothing to do with the blueward turn of the
individual cooling sequences of WDs with H atmospheres, caused by strong
absorption from H$_2$ in the infrared (Hansen 1998, Saumon \& Jacobson
1999).  

That H$_2$-induced turn to the blue sets in only when the $T_{\rm
eff}$ of the models reaches below $\sim 4000$ K, whereas the coolest
temperature reached by the observed WDs is $\sim5400$ K, according to
our 6-Gyr WD isochrone, which matches the faint end of the observed
sequence. 
We also note that some WDs belonging to the brighter peak of the LF
seem to extend farther to the blue than the 4-Gyr isochrone does, and
to even brighter magnitudes.

\subsection{White dwarf cooling times}

In any case, the discrepancies between these WD ages and the age derived
from MS turn-off stars constitute a serious conflict.  There must be some
significant error either in the models that fit the turn-off (which
seems unlikely) or in the models that fit the WD cooling sequence.  We
now look for ways to make our theoretical WDs cool more slowly.

One possible way of slowing the cooling of white dwarfs is to change the
relative abundances of C and O.  The recent Kunz et al.\ (2002) estimate
of the $^{12}$C$(\alpha,\gamma)^{16}$O reaction rate is only about 2/3
of the Caughlan \& Fowler~(1985) rate employed to compute the
evolutionary CO profiles of the Salaris et al.\ (2000) WD models.  This
lower rate would produce a C/O ratio in the core of WDs that is closer
to unity compared with our adopted models.  This would, in turn,
increase the delay introduced by the CO separation upon crystallization.
Taking advantage of the analyses of Montgomery et al.\ (1999) and Isern
et al.\ (2000), we estimate that the extra delay in the cooling process
would be in any case less than 1~Gyr at the luminosities corresponding
to the two observed peaks ($\log(L/L_{\odot})\sim -3.7$ and $\sim
-$4.0).

An additional effect that could slow down the cooling process is the
diffusion of $^{22}$Ne in the liquid phase.  The Ne abundance should
be about 4\% by mass, and its diffusion is not included in any of the
existing large WD model grids.  Calculations by Deloye \& Bildsten
(2002) show that the delays produced by this process are $\sim$ 1~Gyr
for the NGC 6791 WDs.  Very recent calculations by Garcia-Berro et
al.\ (2007) do not change this picture appreciably, if one makes the
most realistic assumptions about the $^{22}$Ne diffusion coefficient,
carbon and oxygen stratification, and crystallization process.

A third process that could potentially add an extra delay to the WD
cooling timescales is the $^{22}$Ne separation during the
crystallization phase (Segretain et al.\ 1994). Our adopted WD models
include only the effect of separation of a CO binary mixture, without
taking into account the presence of a small amount of $^{22}$Ne.
Segretain~(1996) computed the phase diagram of a ternary C/O/Ne
mixture, and estimated an extra delay of $\sim0.4$~Gyr due to Ne
separation, for a 1\% Ne mass fraction in the WD core.  This delay
would affect the theoretical WD isochrones and LFs only marginally;
however, one needs computations of this effect for a larger Ne
abundance typical of NGC 6791 WDs, and more realistic CO profiles
(Segretain~1996 consider a flat C and O profile, with 49.5\% mass
fraction each), before drawing definitive conclusions about the
relevance of this process.

The effect of one or more of these processes might allow isochrones of a
reasonable age to fit the fainter peak, but it is very difficult, if not
impossible, to reproduce the brighter one.  In any case, we would be
ignoring the glaring problem of having two distinct peaks in the WDLF.

\subsection{Helium white dwarfs}

An obvious way of dealing with the two peaks in the WD LF is to assume
that the
cluster has two different kinds of white dwarf, and a reasonable
hypothesis would be that in addition to the usual CO white dwarfs the
cluster has a population of white dwarfs whose
cores consist of helium.

In fact, Hansen (2005) has put forward the idea that the bright peak
at $m\subr{F606W} \sim 27.4$ is caused by massive helium WDs, produced
by RGB stars that just missed He ignition (because of mass loss due to
stellar winds).
This idea is supported by the fact that NGC 6791 contains a 
non-negligible number of blue He-burning stars that have very little
mass in their envelopes, having lost nearly all of the envelope during
their RGB lives.
Kalirai et al.\ (2007) have recently estimated the masses of several
objects along the bright part of the WD sequence and conclude that a
substantial fraction of the NGC 6791 WDs are actually helium WDs.
Given that He WDs cool down more slowly than CO WDs, an ad hoc choice
of the relationship between the mass (and number) of RGB stars that
have lost all their envelope before the He-ignition, on the one hand,
and the final He WD mass, on the other hand, can produce a LF that
reproduces the bright observed peak for an age compatible with the TO
age (Hansen 2005).

\begin{figure}
\epsscale{1.00}
\plotone{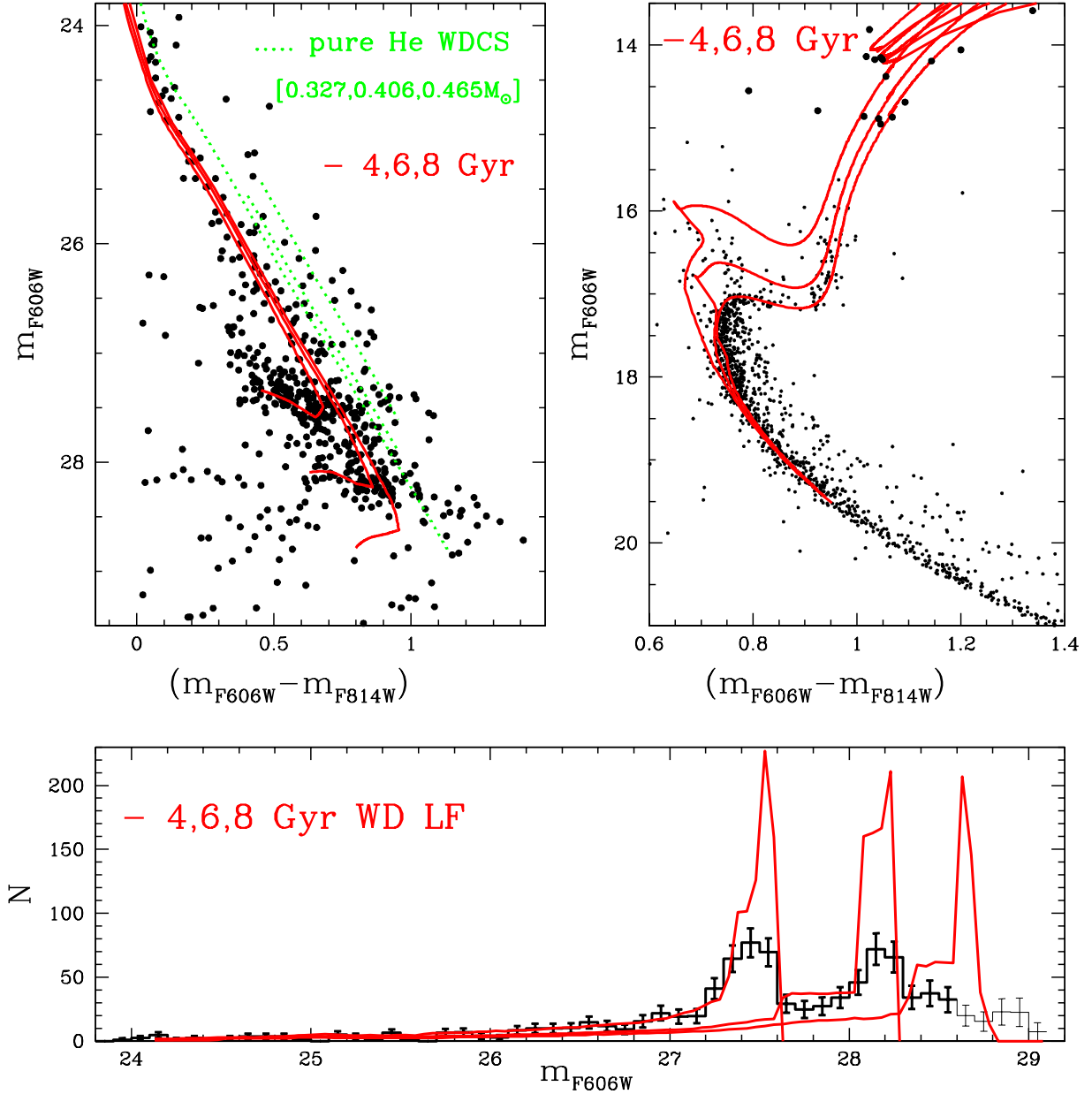}
\caption{Fit of theoretical isochrones and LFs of CO WDs for ages 4, 6,
         and 8~Gyr to the observed CMD of NGC 6791 and to the WD star
         counts (see text for details).  Auxiliary quantities used here
         are $E(B-V)=0.17$, $(m-M)_V=13.5$, $R_V=3.2$,
         $A\subr{F606W}=0.47$, $A\subr{F814W}=0.31$. Cooling sequences
         of He WDs with masses equal to, respectively, 0.327, 0.406, and
         0.465$M_{\odot}$ are also displayed (the bluest being the most
         massive).  Note that no normalization of the theoretical LFs to
         the observed numbers has been attempted, as the only purpose of
         the comparison in the lower panel is to show the expected
         position in magnitude of the peaks for the different assumed
         ages.}
\label{fit}
\end{figure}

A serious problem with this scenario is that the He-core mass at the
He flash is 0.46$M_{\odot}$ at the super-solar metallicity of NGC 6791
(Pietrinferni et al.\ 2004).
Even at solar metallicity it is just 0.47$M_{\odot}$. We have therefore
computed the evolution of a 0.465$M_{\odot}$ He WD using the code by
Serenelli et al.\ (2002). The color transformations applied to this
model are the same as for the CO WD isochrones.  The top-left panel of
Fig.~\ref{fit} displays this sequence (as a green dotted line) together
with additional He WD models of lower mass, already shown in Bedin et
al.~(2005a).  These low-mass models overlap with objects redder than the
main body of the cooling sequence, which most likely correspond to the
fainter counterpart of the bright He WDs identified by Kalirai et
al.~(2007).  As for the 0.465$M_{\odot}$ track, this is too red by
$\sim0.08$ mag in $(m_{\rm F606W}-m_{\rm F814W})$ compared to the main
branch of the CO isochrone that reproduces very well the main body of
the observed cooling sequence.  One would need He WDs above $\sim
0.5~M_{\odot}$ to eliminate this discrepancy.  Another important issue
is whether He-core WDs are able to reproduce the objects along the blue
hook corresponding to the bright peak in the LF.  We stress again that
the $T_{\rm eff}$ of both CO- and He-core WDs at these magnitudes are
both well above 4000~K; therefore one cannot invoke the turn to the blue
caused by H$_2$ absorption (which in any case is included in our adopted
color transformations) to explain the observed blue hook in terms of
single-mass cooling sequences.  Thus for the case of He WDs this feature
can be caused only by the presence of a range of masses that has to
extend above $0.5~M_{\odot}$.

As an extreme test, one could in principle disregard the fit to the main
sequence, the TO and the red giant branch, and consider only the
observed WD sequence. Assuming $E(B-V)=0.09$--0.10, the lowest reddening
among the values quoted in the literature (see Stetson et al.\ 2003),
the 0.465$M_{\odot}$ He WD track overlaps with the main body of the
cooling sequence --- with this $E(B-V)$ value, main sequence and red
giant branch would not be matched by theoretical models --- but
obviously fails to match the blue hook.  To reproduce this feature
(leaving aside the resulting values of the cooling age and the TO age)
one has to unrealistically increase the cluster distance modulus by
$\sim$ 1~mag --- i.e., $(m-M)_V \sim 14.5$, which corresponds to
$(m-M)_0 \sim$ 14.2 for $E(B-V)=0.09$--0.10 --- so that the
0.465$M_{\odot}$ model would intersect the bluest point along the blue
hook, and lower masses would progressively shift towards the red side of
the hook.  Even before considering the adequacy of the resulting cooling
ages, one has to note that such a large distance modulus is well outside
the range of all reasonable determinations (Stetson et al.\ 2003). Again
the conclusion is that He-core WDs with a spectrum of masses extending
well beyond 0.5$M_{\odot}$ are necessary if one wants to explain the
observed bright peak of the LF in terms of these objects.
 
One physical mechanism that could produce massive He WDs is rotation.
Mengel \& Gross (1976) showed that rotation can increase the value of
the core mass at the He flash by as much as 0.15$M_{\odot}$ of a solar mass,
compared with standard non-rotating models.  
As a general rule, when the angular-momentum transport mechanism is
fixed, faster-rotating stars will reach larger core masses at the time
of He flash.  As a by-product, according to the Reimers (1975)
mass-loss law, rotating stellar models lose more mass when approaching
the He flash compared with non-rotating or slowly rotating ones. This
provides a natural mechanism that favors the production of blue
He-burning objects (extreme blue HB, as observed in this cluster) and
massive He WDs (which might explain the brighter peak of the WDLF).
Stellar rotation causes a broadening of the lower part of the RGB
(Brown 2007).  Curiously enough, this broadening of the base of the
RGB (broader than expected from observational errors) seems to be
present in the CMDs from Stetson et al.\ (2003, see their Fig.~17),
Bruntt et al.\ (2003, see their Figs. 2, 3, 5, and 10), and in our own
CMD.  Note that this RGB broadening cannot be due to a spread in
metallicity, as Carraro et al.\ (2006) find a negligible ($\pm0.01$
dex) dispersion in [Fe/H].  In addition, extra mixing induced by
rotation of the stars (Charbonnel 1995, and references therein) leads
to a low value of $^{12}$C/$^{13}$C, as observed by Origlia et al.\
(2006) in RGB stars of NGC 6791.

We are currently investigating (Brown et al., in preparation) 
what initial rotation rates and angular-momentum transport prescriptions are 
necessary to produce He-core masses $\sim0.5M_{\odot}$ and higher at 
the metallicity of NGC 6791, the impact of rotation on the theoretical 
CMD, and the corresponding predicted surface rotation velocities along 
the CMD. This investigation, together with spectroscopic measurements 
(presently lacking) of surface rotation of MS and TO stars may definitively 
prove or disprove this scenario.

\acknowledgements
J.A.\ and I.R.K.\ acknowledge support from STScI grants GO-9815 and 
GO-10471. 
We thank Alvio Renzini for many useful discussions.
We warmly thank Lars Bildsten and Christopher Deloye for valuable
discussions about $^{22}$Ne.  We also thank the anonymous referee for
the careful reading and for useful suggestions.

\end{document}